\documentclass[reprint,
superscriptaddress,
 amsmath,amssymb,
 aps,
]{revtex4-1}

\usepackage{graphicx}% Include figure files
\usepackage{dcolumn}% Align table columns on decimal point
\usepackage{bm}% bold math
\usepackage[utf8]{inputenc}
\usepackage[T1]{fontenc}
\usepackage{mathptmx}
\usepackage{amsmath,amsfonts,amssymb}
\usepackage{graphicx}
\usepackage{setspace}
\usepackage{tocloft}
\usepackage{siunitx}
\usepackage[version=4]{mhchem}
\begin{document}
\title[Non-thermal light-assisted resistance collapse in a V$_2$O$_3$-based Mottronic device]{Non-thermal light-assisted resistance collapse in a V$_2$O$_3$-based Mottronic device}
% Force line breaks with \\

\author{A. Ronchi}%
 \email{andrea.ronchi@unicatt.it.}
\affiliation{Department of Physics and Astronomy, KU Leuven, Celestijnenlaan 200D, 3001 Leuven, Belgium}
\affiliation{Department of Mathematics and Physics, Universit\`a Cattolica del Sacro Cuore, Brescia I-25121, Italy}
\affiliation{ILAMP (Interdisciplinary Laboratories for Advanced Materials Physics), Universit\`a Cattolica del Sacro Cuore, Brescia I-25121, Italy}

\author{P. Franceschini}
\affiliation{Department of Physics and Astronomy, KU Leuven, Celestijnenlaan 200D, 3001 Leuven, Belgium}
\affiliation{Department of Mathematics and Physics, Universit\`a Cattolica del Sacro Cuore, Brescia I-25121, Italy}
\affiliation{ILAMP (Interdisciplinary Laboratories for Advanced Materials Physics), Universit\`a Cattolica del Sacro Cuore, Brescia I-25121, Italy}

\author{P. Homm}
\affiliation{Department of Physics and Astronomy, KU Leuven, Celestijnenlaan 200D, 3001 Leuven, Belgium}

\author{M. Gandolfi}
\affiliation{CNR-INO, Via Branze 45, 25123 Brescia, Italy}
\affiliation{Department of Information Engineering, University of Brescia, Via Branze 38, 25123 Brescia, Italy}

\author{G. Ferrini}
\affiliation{Department of Mathematics and Physics, Universit\`a Cattolica del Sacro Cuore, Brescia I-25121, Italy}
\affiliation{ILAMP (Interdisciplinary Laboratories for Advanced Materials Physics), Universit\`a Cattolica del Sacro Cuore, Brescia I-25121, Italy}

\author{S. Pagliara}
\affiliation{Department of Mathematics and Physics, Universit\`a Cattolica del Sacro Cuore, Brescia I-25121, Italy}
\affiliation{ILAMP (Interdisciplinary Laboratories for Advanced Materials Physics), Universit\`a Cattolica del Sacro Cuore, Brescia I-25121, Italy}

\author{F. Banfi}
\affiliation{Universit\'e de Lyon, Institut Lumière Mati\'ere (iLM), Universit\'e Lyon 1 and CNRS, 10 rue Ada Byron, 69622 Villeurbanne Cedex, France}

\author{M. Menghini}
\affiliation{IMDEA Nanociencia, Cantoblanco, 28049 Madrid, Spain}
\affiliation{Department of Physics and Astronomy, KU Leuven, Celestijnenlaan 200D, 3001 Leuven, Belgium}

\author{J-.P. Locquet}
\affiliation{Department of Physics and Astronomy, KU Leuven, Celestijnenlaan 200D, 3001 Leuven, Belgium}

\author{C. Giannetti}
    \email{claudio.giannetti@unicatt.it.}
\affiliation{Department of Mathematics and Physics, Universit\`a Cattolica del Sacro Cuore, Brescia I-25121, Italy}
\affiliation{ILAMP (Interdisciplinary Laboratories for Advanced Materials Physics), Universit\`a Cattolica del Sacro Cuore, Brescia I-25121, Italy}

\date{\today}% It is always \today, today,

\begin{abstract}
The insulator-to-metal transition in Mott insulators is the key mechanism for a novel class of electronic devices, belonging to the Mottronics family. Intense research efforts are currently devoted to the development of specific control protocols, usually based on the application of voltage, strain, pressure and light excitation. The ultimate goal is to achieve the complete control of the electronic phase transformation, with dramatic impact on the performance, for example, of resistive switching devices. Here, we investigate the simultaneous effect of external voltage and excitation by ultrashort light pulses on a single Mottronic device based on a V$_2$O$_3$ epitaxial thin film. The experimental results, supported by finite-element simulations of the thermal problem, demonstrate that the combination of light excitation and external electrical bias drives a volatile resistivity drop which goes beyond the combined effect of laser and Joule heating. Our results impact on the development of protocols for the non-thermal control of the resistive switching transition in correlated materials.
\end{abstract}

\maketitle

\section{\label{sec:intro} Introduction}
Mott insulators are a class of quantum materials exhibiting the most promising properties and functionalities for the next generation of solid-state devices belonging to the Mottronics family\cite{Tokura2017,Yang2011,Zhang2014,Basov2017}. Correlated vanadium-oxides have been intensively studied due to the possibility of selectively controlling the resistive switching process at the core of resitive memories and neuromorphic computing\cite{Zhou2015,Janod2015,delValle2018,Salev2019}. The conventional route to control the resistive switching dynamics is based on the tuning of chemical doping\cite{Kuwamoto1980,Homm2015,delvalle2017}, temperature \cite{Ronchi2019,Mcleod2016}, pressure\cite{McWhan1969,Jayaraman1970,Valmianski2017,Limelette89}, strain\cite{Dillemans2014,Lee2018,Kalcheim2019} and on the application of external electric fields\cite{Stoliar2013,Guenon2013,Mazza2016,DelValle2019, Kalcheim2020}. More recently, the sudden excitation via ultrashort laser pulses\cite{Mansart2010,Liu2011,Abreu2015,Morrison2014,Abreu2017,Lantz2017,Ronchi2019,Otto2019,Giorgianni2019} has been introduced as a novel control parameter, although the nature of the photoinduced insulator-to-metal transition is still subject of intense debate. Much effort has been so far dedicated to the study of the transient optical, electronic and lattice properties\cite{Mansart2010,Liu2011,Abreu2015,Morrison2014,Abreu2017,Lantz2017,Ronchi2019,Otto2019,Giorgianni2019} and on the role of intrinsic nanoscale inhomogeneities\cite{Lupi2010,Mcleod2016,Ronchi2019} with the goal of understanding to what extent the photoinduced transition is similar to the thermally driven one. In contrast, little is known about the actual resistive state that is induced by light excitation possibly combined with external voltage. Recent theoretical works have suggested that, in the vicinity of the insulator-metal coexisting region, the application of voltage\cite{Mazza2016} and the excitation with light pulses capable of manipulating the band occupation\cite{Sandri2015,Ronchi2019}, can lead to the weakening of the insulating state and, eventually, to the complete collapse of the Mott insulating phase and the consequent resistive switching along a non-thermal pathway.

Here we developed a Mottronic device, based on a micro-bridged V$_{2}$O$_{3}$ epitaxial thin film, in which the resistance state consequent to a controlled excitation by trains of ultrashort light pulses can be measured. V$_{2}$O$_{3}$ is a prototypical Mott insulator that exhibits a first-order Insulator-to-Metal transition (IMT) occurring at $T_{IMT}$ $\sim$ 170 \si{\kelvin} and characterized by a resistivity change of several orders of magnitude\cite{McWhan1969,Jayaraman1970,Limelette89}. When a voltage larger than a temperature-dependent threshold $\Delta V_{th}$ is applied in the insulating state ($T<T_{IMT}$), the device undergoes a resistive switching process, i.e. the resistance suddenly drops to the metallic value. However, the switching process in V$_{2}$O$_{3}$ devices is associated to the formation of micrometric conductive filaments\cite{Kalcheim2020}, which usually makes the evaluation of the local heating and the investigation of the thermal nature of the dynamics rather difficult. In order to avoid this additional complexity, we investigate the light induced volatile resistance drop for different applied voltages, always smaller than $\Delta V_{th}$. This below-threshold  resistance drop thus provides information about the weakening of the insulating phase, triggered by the combined voltage-light excitation, while avoiding the formation of the conductive channels associated to the resistive switching process. Starting from the resistance vs temperature 
calibration curve of the device, we are able to compare the photoinduced resistivity drop to the local heating effect, which is carefully estimated by finite-element simulations. Our main results can be summarized as follows. The application of a sufficiently long train of light pulses alone drives a volatile resistance drop that is compatible with that predicted by finite-element simulations of the thermal problem. The minimum number of pulses necessary to observe a significant effect suggests that the light-induced local heating is mediated by the electronic specific heat, which becomes extremely large as $T_{IMT}$ is approached. Furthermore, when a significant below-threshold electrical bias ($\Delta V$=0.5 V) is applied, the light-induced volatile drop of the resistance is twice than what is expected by considering both the laser-induced and Joule heating. This result leads to the conclusion that the combined voltage-light excitation protocol makes the system more fragile towards the collapse of the insulating electronic phase.

%In particular, we will focus on vanadium sesquioxide (V$_{2}$O$_{3}$), a prototypical Mott insulator that exhibits a Insulator-to-Metal transition (IMT) occurring at T$_{IMT}$ $\sim$ 160 \si{\kelvin} and characterized by a resistivity change of several orders of magnitude. Despite this, the microscopic mechanism driving the IMT in this material is still the subject of a lively debate and recent results unveiled a complex interplay between electronic correlations [\cite{Imada1998}], lattice deformation [\cite{Singer2018,Dernier1970}] and intrinsic inhomogeneities at the nanoscale [\cite{Ronchi2019,Lupi2010,Lantz2015,Mcleod2016}]. In particular, investigating the way this inhomogeneities rule the dynamics of the IMT transition is the keypoint for looking into the possibility of properly engineering them such to expect an ultrafast control of the transition. Within this scenario, resisitivity measurements are a direct probe of the interconnection between insulating and metallic domains at the micro/nanoscale since the resistive transition occurs when a connected filamentary path of metallized nodes forms [\cite{Stoliar2013}].\par
%Here we combine femtosecond laser pulses excitation with the resistivity measurements to investigate the laser-induced thermal heating of a V$_{2}$O$_{3}$ micro bridge. The experimental results are accompanied by Finite Element Method (FEM) simulations.
%%%%%%%%%%%%%%%%%%%%%%%%%%%%%%%%%%%%%%%%%%%%%%%%%%%%%%%%%%%%%%%%%%%%
\section{Methods\label{materials}}
\subsection{Mottronic device}
\label{sec:device}
An epitaxial V$_{2}$O$_{3}$ film with thickness $d$=67 nm is deposited by oxygen-assisted molecular beam epitaxy (MBE) in a vacuum chamber with a base pressure of $10^{-9}$ Torr. A 37 \si{nm} Cr$_{2}$O$_{3}$ buffer layer is interposed between the film and the substrate to minimise lattice mismatch and optimise strain relaxation in the film.
A (0001)-Al$_{2}$O$_{3}$ substrate is used without prior cleaning and is slowly heated to a growth temperature of 700 \si{\celsius}. Vanadium is evaporated from an electron gun with a deposition rate of 0.1 \si{\angstrom \per \second}, and an oxygen partial pressure of $6.2\cdot 10^{-6}$ Torr is used during the growth \cite{Dillemans2014}. Under these conditions, a single crystalline film with the $c$-axis oriented perpendicular to the surface is obtained.
A micro bridge, constituted by 40 \si{nm} Au/5 \si{nm} Ti thick electrodes with $w$=50 \si{\mu\metre} width and $s$=2 \si{\mu\metre} separation, is nanopatterned on the film surface, as shown in Fig. \ref{fig:1}a. Temperature-dependent resistivity measurements (see Fig. \ref{fig:1}b) are performed using a Keithley Sourcemeter 2634B in the 2-points configuration across the Au/Ti electrodes. The device temperature is controlled by a closed cycle cryostat equipped with a heater. The temperature sweep rate is set to 0.5 \si{\kelvin} per minute. When cooled down, the device undergoes the metal-to-insulator transition, characterized by a resistance increase of almost three order of magnitudes. In our device  the resistance ranges from 20 \si{\Omega} at room temperature (metallic state) to 130 \si{k \Omega} at $T$=140 K (insulating state). The heating and cooling branches evidence the typical hysteresis cycle of the IMT, which spans the insulator/metal coexistence region from 155 to 185 K. In the following, we will consider $T_{IMT}\simeq$170 K as the temperature corresponding to the midpoint of the heating branch of the hysteresis cycle.  
%%%%%%%%%%%%%%%%%%%%%%%%%%%%%%%%%%%w

\subsection{Photo-induced resistance drop measurements}
\label{sec:PI_res_measurements}
\begin{figure}
\begin{center}
\begin{tabular}{c}
\includegraphics[width=8.6cm]{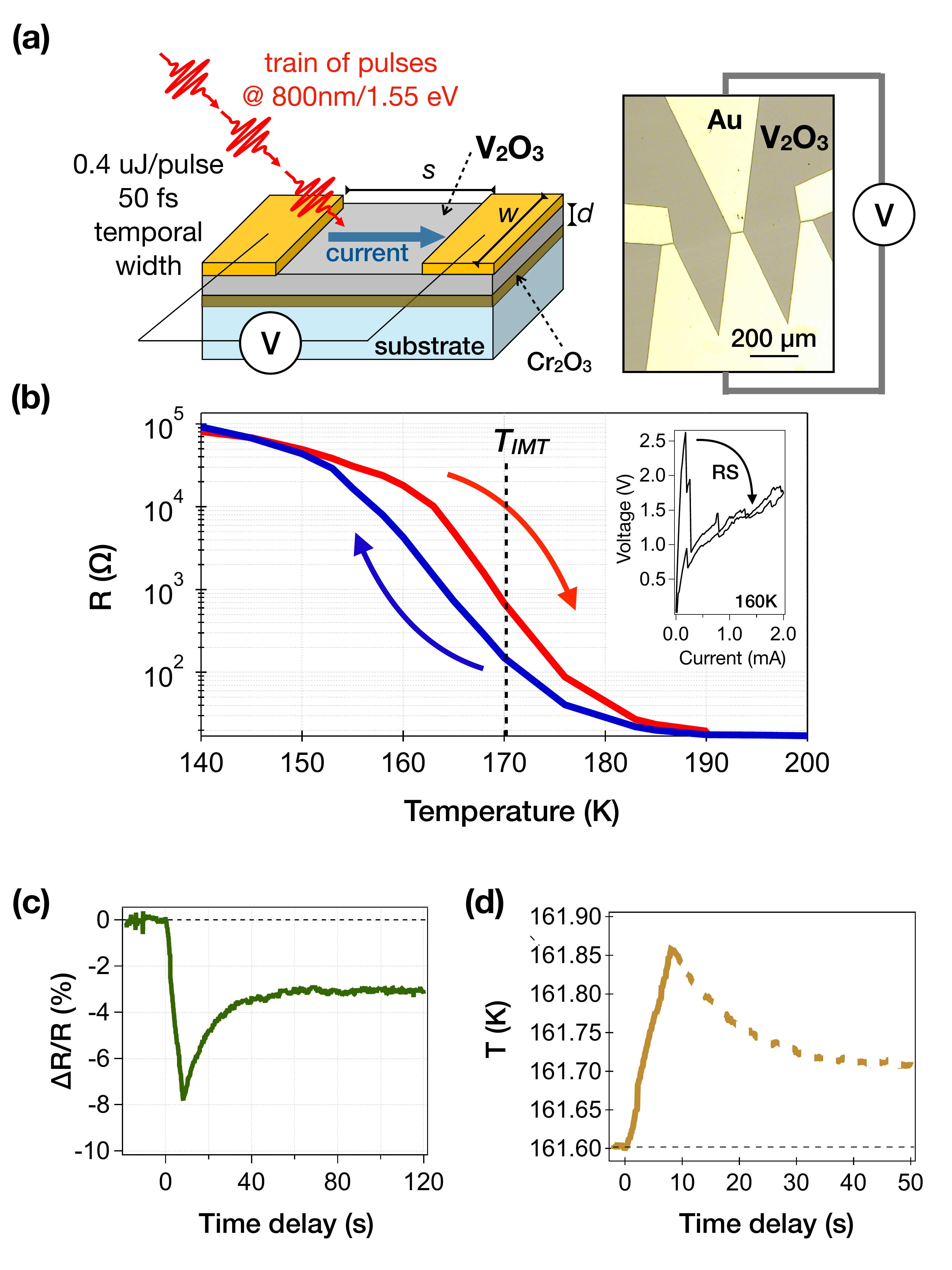}
\end{tabular}
\end{center}
\caption 
{ \label{fig:1}
\textbf{Working principle of the experiment.} \textbf{(a)} Left: Cartoon of the experimental set-up. A finite train of laser pulses ($\hbar\omega=1.55$ \si{\eV}, $50$ \si{fs} temporal width, $\nu=25$ \si{kHz}) is used to photoexcite the device. The change in the 2-points resistance of the device is simultaneously measured before, during and after the laser excitation by means of Au/Ti electrodes. The contact resistance can be neglected since is very small ($\sim$ 5 \si{\ohm}) with respect to the V$_{2}$O$_{3}$ bridge resistance ($\sim$ 15 \si{k\ohm}). Right: Microscopic image of the device used during the experiment. \textbf{(b)} Equilibrium 2-points resistance hysteresis measured as function of the temperature. The red (blue) arrows indicate the heating (cooling) branch. Inset: 2-points I-V curve showing volatile resistive switching at $T$=160 K. The measurement has been performed through the 2 \si{\mu m} gap shown in (a). \textbf{(c)} Relative variation of resistance measured across the device as a function of the time delay at the base temperature $T_0$=161.6 K. The time zero corresponds to the moment when the first pulse of the laser train arrives. \textbf{(d)} Result of the conversion of resistance drop measurements (see panel (c)) into effective local temperature (yellow solid line). The resistance vs temperature curve shown in Fig.\ref{fig:1}(b) has been inverted to obtain, for each value of resistance, the corresponding local temperature. As explained in the main text, the cooling down process takes place along a resistance-temperature curve which differs from the equilibrium one reported in panel (b). The dashed line, corresponding to the cooling down dynamics of the system, is the hypothetical local temperature obtained from the heating branch of the equilibrium resistance-temperature curve, but does not necessarily represents the actual local effective temperature of the system.}
\end{figure}

Resistance measurements are combined with light excitation by focusing a train of infrared pulses (\text0.4 \si{\mu \joule}, 50 \si{fs}) at 800 \si{nm} wavelength (1.55 eV photon energy) generated by an optical parametric amplifier pumped by an Yb-laser system. A pulse picker allows to control both the repetition rate, set to 25 \si{kHz}, and the total number of pulses impinging on the sample. The laser output is focused on the device by a 10 \si{cm} lens. The Full-Width-Half-Maximum (FWHM) of the spot size on the device is 50 \si{\mu \meter}, as measured using the knife-edge technique. The pump incident fluence on the device is of the order of 0.2 mJ/cm$^2$, which is below the threshold necessary to photoinduce the complete insulator-to-metal transition\cite{Ronchi2019}. In Fig. \ref{fig:1}c we show the typical resistance drop measured after excitation with a controlled number of light pulses (200 kpulses at 25 kHz rep. rate). Prior to the experiment the device is cooled down to 100 K and then heated up to $T\simeq$160 K, i.e. in the insulator-metal coexistence region. Before light excitation, the system is thus in a high resistivity state along the heating branch of the hysteresis curve shown in Fig. \ref{fig:1}b. Laser excitation is then switched on and the time-dependent resistance is obtained by measuring the current flowing across the bridge at a fixed applied voltage ($\Delta V$=5 mV) and at the nominal temperature $T$=160 K. After 8 s, i.e. at the end of the pulse train, the resistance drops by 8 \% with respect to the initial value. Once the light excitation is removed, the relative resistance variation, i.e. $\Delta R/R$, starts to decrease, until it reaches a plateau corresponding to a non-volatile change of about 3\%. The explanation of this effect resides in the inherent hysteresis of the IMT (see Fig.\ref{fig:1} b): after the laser-induced warming up, which is not enough to drive the system out of the hysteresis region, the cooling-down takes place along a curve located somewhere between the complete warming up and cooling down curves (red and blue curves in Fig. \ref{fig:1}b). The final resistance value is therefore smaller than the initial value, although the initial temperature is recovered. In order to restore the initial resistance value, the device must undergo a complete thermal cycle, corresponding to heating up to 300 K, followed by cooling down to 100 K and further heating up to $T\simeq$160 K. The same thermal cycle protocol is applied before any of the resistance drop measurements reported in the following.

The resistance-temperature curve shown in Fig.\ref{fig:1}b can be used to retrieve the precise local effective temperature of the V$_2$O$_3$ device during the light excitation process, as shown in Fig. \ref{fig:1}d. The local effective temperature during the resistance drop is extracted by interpolating and inverting the heating branch of the resistance-temperature curve (Fig. \ref{fig:1}b). %A twelfth degree polynomial function is fit to the curve obtained by the inversion procedure. The fit coefficients are used in an automatic routine to convert each point of all the experimental curves in local effective temperature (see Fig. \ref{fig:1}d). 
The error associated to the local effective temperature is obtained by considering the accuracy of the Keithley Sourcemeter ($\pm$ 0.02 \%) in the voltage range used during the experiments. We stress that the resistance-temperature relation is single-valued only along the heating branch. Therefore, the resistance value can be converted in the actual local temperature only during the heating up process in the 0-8 \si{s} time span. During the cooling down (8-120 ps) the resistance moves back along a non-equilibrium resistance-temperature curve, which differs from that used for calibration (Fig. \ref{fig:1}b) and cannot therefore be used to retrieve the actual local temperature.

\subsection{Finite Elements Method simulation of the thermal problem}
\label{sec:thermal}
\begin{figure}
\begin{center}
\begin{tabular}{c}
\includegraphics[width=8.2cm]{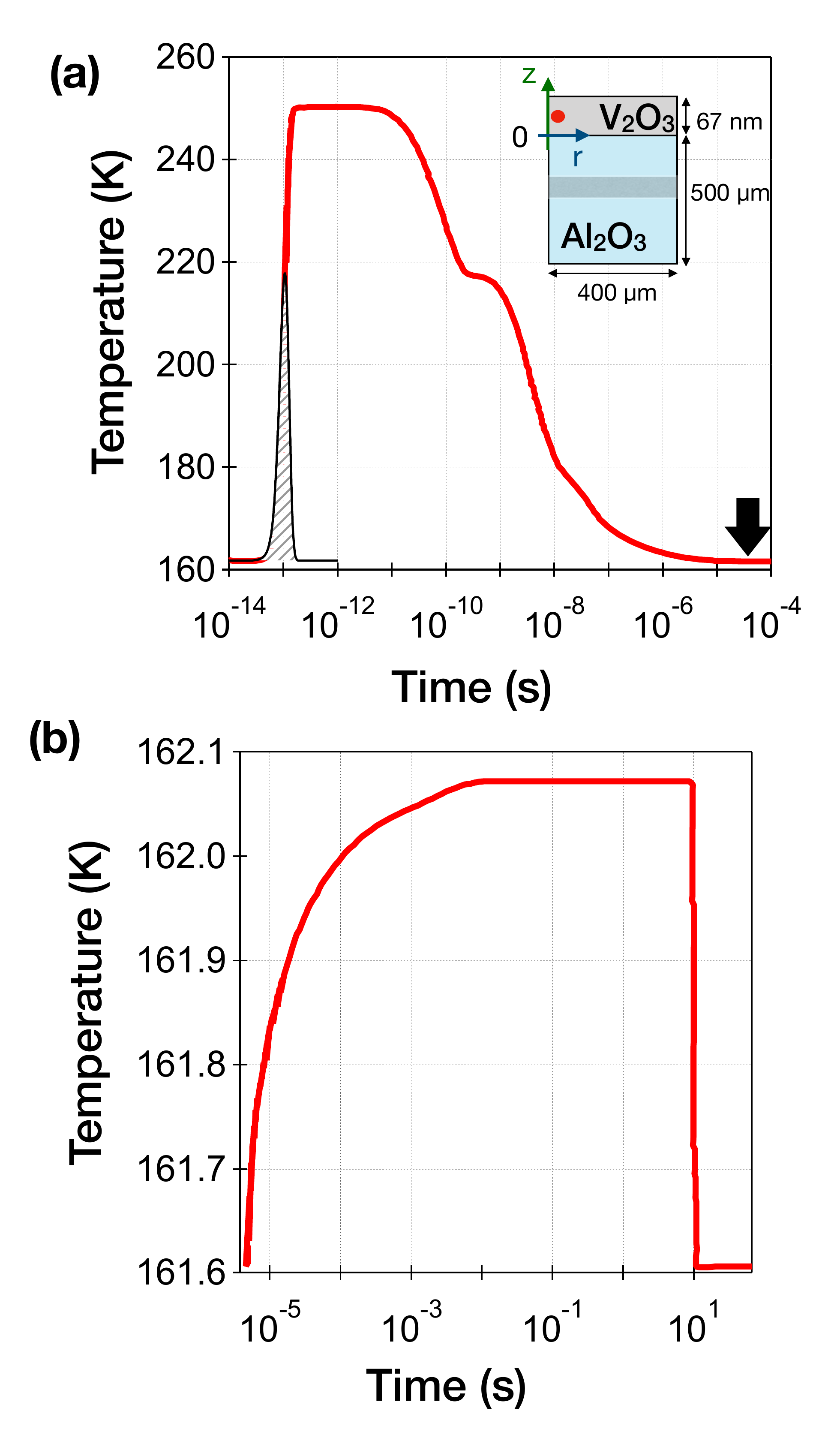}
\end{tabular}
\end{center}
\caption 
{ \label{fig:2}
\textbf{Finite elements method simulations.} \textbf{(a)} The solid line represents the temperature increase due to the single pulse excitation (black dashed area) as function of the time (logarithmic scale). The black arrow indicates the instant ($t=40$ \si{\mu\second}) at which the following pulse arrives. Inset: sketch of the geometry used in the Comsol simulations. The red dot indicates the point at which the temperature has been evaluated ($r=50$ \si{nm}; $z$=33.5 \si{nm}). \textbf{(b)} The temperature variation at saturation, induced by $2\cdot10^{5}$ laser pulses, is shown. The temperature is evaluated at half of the total thickness, i.e. $r=50$ \si{nm} and $z$=33.5 \si{nm}. The maximum temperature reached after the excitation with the pulse train is used as input to build the curves reported in Fig. \ref{fig:3}. } 
\end{figure}

The heating induced by the laser excitation is numerically simulated by finite-element methods (Comsol Multiphysics). In the simulations we consider the 67 \si{nm} thick V$_{2}$O$_{3}$ film on top of a 500 \si{\mu \metre} thick (0001)-Al$_{2}$O$_{3}$ substrate. The Cr$_{2}$O$_{3}$ buffer-layer is omitted since its thermal and optical properties are very similar to the sapphire's ones. The problem is solved in 2D axial symmetry in a region with 800 \si{\mu\meter} diameter. A representation of the sample cross section is reported in the inset of Fig. \ref{fig:2}a.
Thermal insulating boundary conditions are applied on the top and lateral boundaries, whereas the sapphire bottom boundary ($z =-500$ \si{\mu\meter}) is kept at the constant temperature $T_{0}=161.6$ \si{\kelvin}, which corresponds to the actual initial effective temperature of the device, as extracted from the resistance value. 

The thermal problem for single pulse excitation is initially addressed. In this configuration, a single pulse with temporal and spatial gaussian profiles matching the experimental ones, excites the sample. We consider an energy density delivered to the system (0.66 \si{kJ/cm^3}) such that the time and spatial integral matches the total energy released by the laser pulse upon accounting for reflection losses. The estimated energy absorbed  within the 67 \si{nm} V$_{2}$O$_{3}$ film is calculated with a multireflection model (total absorption $A$= 0.22), assuming the optical constants reported in the literature\cite{Qazilbash2008}. 
Using Fourier's law, the energy balance within the volume of the V$_{2}$O$_{3}$ film reads:
\begin{multline}
-\nabla\cdot \bigl[ -\kappa(T(r,z,t))\nabla T(r,z,t)\big]+\\
+P(r,z,t) =\rho\ c_{p}((r,z,t))\frac{\partial T(r,z,t)}{\partial t}
\label{eq:1}
\end{multline}
where $P(r,z,t)$ is the absorbed pulsed power density profile, $\kappa(T)$ is the temperature dependent thermal conductivity\cite{Andreev1978}, $\rho$ the mass density, and $c_{p}(T)$ the temperature-dependent total specific heat at constant pressure\cite{Keer1976}. For sake of completeness, the parameters used in the simulation are reported in Table \ref{tab:1}. 
\begin{table}[t]
\caption{Summary of materials parameters values adopted in the present work for FEM simulations.} 
\label{tab:1}
\begin{center}       
\begin{tabular}{cccc} %% this creates two columns
%% |l|l| to left justify each column entry
%% |c|c| to center each column entry
%% use of \rule[]{}{} below opens up each row
\hline\hline
\rule[-1ex]{0pt}{3.5ex}  \textbf{Parameter}  & \textbf{Value} & \textbf{Units} & \textbf{Ref.}  \\
\hline
\rule[-1ex]{0pt}{3.5ex} $c_p(T)\big|_{V_2O_3}$  & $c_p$($T$=160 K) =480 & \si{\joule\ \kilogram^{-1}\ \kelvin^{-1}} & \cite{Keer1976} \\
\rule[-1ex]{0pt}{3.5ex}  & $c_p$($T_{IMT}$)=2400 & \si{\joule\ \kilogram^{-1}\ \kelvin^{-1}} & \cite{Keer1976}
\\
\rule[-1ex]{0pt}{3.5ex}  & $c_p$($T$=190 K)=530 & \si{\joule\ \kilogram^{-1}\ \kelvin^{-1}} & \cite{Keer1976}\\
\hline
\rule[-1ex]{0pt}{3.5ex} $c_p \big|_{Al_2O_3}$  & 776 & \si{\joule\ \kilogram^{-1}\ \kelvin^{-1}} & \cite{Weber2001} \\
\hline
\rule[-1ex]{0pt}{3.5ex} $\kappa(T)\big|_{V_2O_3}$  & $\kappa(T)$ ($\sim$3.1 @160 \si{\kelvin}) & \si{W\ \metre^{-1}\ \kelvin^{-1}} & \cite{Andreev1978} \\
\hline
\rule[-1ex]{0pt}{3.5ex} $\kappa\big|_{Al_2O_3}$  & 35 & \si{W\ \metre^{-1}\ \kelvin^{-1}}  & \cite{Caddeo2017} \\
\hline
\rule[-1ex]{0pt}{3.5ex} $\rho_{V_{2}O_{3}} $ & 4870 & \si{\kilogram\ \metre^{-3}}  & \cite{Weber2001} \\
\hline
\rule[-1ex]{0pt}{3.5ex} $\rho_{Al_2O_3}$  & 3980 & \si{\kilogram\ \metre^{-3}} & \cite{Weber2001} \\
\hline
\rule[-1ex]{0pt}{3.5ex} Re($\tilde{n}_{V_2O_3}$) & 1.997 @800 nm & \si{\kilogram\ \metre^{-3}} & \cite{Qazilbash2008} \\
\hline
\rule[-1ex]{0pt}{3.5ex} Im($\tilde{n}_{V_2O_3}$)  & 0.382 @800 nm & \si{\kilogram\ \metre^{-3}} & \cite{Qazilbash2008} \\
\hline
\rule[-1ex]{0pt}{3.5ex} Re($\tilde{n}_{Al_2O_3}$)  & 1.76 @800 nm& \si{\kilogram\ \metre^{-3}} & \cite{Malitson1962} \\
\hline\hline
\end{tabular}
\end{center}
\end{table} 
The following heat flux continuity boundary condition is used at the V$_2$O$_3$/ Al$_2$O$_3$ interface:
\begin{equation}
k_{V_{2}O_{3}}\frac{\partial T}{\partial z}\bigg|_{z=0^{+}}=k_{Al_{2}O_{3}}\frac{\partial T}{\partial z}\bigg|_{z=0^{-}}.
\end{equation}
The initial condition is $T_{0}=161.6$ \si{\kelvin} throughout the sample and the substrate. The equation is solved via the finite element method, covering a time scale spanning seven orders of magnitude. Since the presence of a thermal boundary resistance (TBR) at the V$_2$O$_3$/ Al$_2$O$_3$ interface can affect the temperature of the V$_2$O$_3$ film, simulations with different TBR values were performed. Assuming a TBR (10$^{-7}$ m$^2$KW$^{-1}$) one order of magnitude larger than typical values reported for similar materials\cite{Hamaoui2019}, leads to a correction of the V$_2$O$_3$ temperature profile smaller than 10\%, i.e. within the experimental uncertainty. 
The thermal dynamics of the single pulse excitation is shown in Fig. \ref{fig:2}a. We remark that we are interested in thermal dynamics on timescales longer than the pulse duration and electron-phonon coupling. Therefore, we assume the complete thermalization between the electron and lattice subsystems, as implied by the use of the total specific heat and thermal conductivity values.  

After the impulsive energy absorption, the temperature reaches its maximum value ($\sim$ 250 \si{\kelvin}) within few picoseconds. Subsequently, the energy diffusion through the sapphire substrate drives the slower cooling down process. If we consider the time needed to reduce the initial temperature variation by a factor 2, we obtain a half-life of about 2 ns. The cooling down to the initial temperature $T_{0}$=161.6 K is almost completed within $\sim$ 40 \si{\mu \second}, although a small temperature increase $\sim$ 0.04 \si{K} persists. The simulations also demonstrate that, in the device region, the temperature variation within the V$_2$O$_3$ film is rather homogeneous, with less than 1\% difference between the top and bottom temperatures. For sake of simplicity, in the following we will refer to the temperature of the point at 33.5 nm distance from the device surface (see Fig. \ref{fig:2}a).

In the multi-pulse experimental configuration, the heating associated to each pulse accumulates on top of the temperature increase triggered by the previous ones. Considering a pulse distance of 40 $\mu$s (25 kHz rep. rate), the small temperature difference accumulated after each pulse leads to a progressive increase of the local effective temperature, until the external heat flux is balanced by the dissipation. In this condition, the average effective temperature does not further increase and it reaches a saturation value, which ultimately drives the resistance drop across the bridge observed after the light excitation. For simplicity, the effect of the total thermal heating induced by the multi-pulse excitation is simulated by considering a laser beam with step-like temporal profile extending over the same time span of the pulse train and with the same average power \cite{Giannetti2009,Banfi2010,Caddeo2017}. An example of the temperature dynamics, for a train of $2\cdot 10^{5}$ pulses, is shown in Fig. \ref{fig:2}b. The heat accumulation leads, at saturation, to a final temperature of 162.1 \si{\kelvin}. The discrepancy between this value and the experimental one (see Fig. \ref{fig:2}b) will be clarified in the next section. Once the input power is removed, the system rapidly cools down to the initial temperature. For a better comparison with the experimental data, the time-dependent temperature curves obtained by the simulations and reported in Fig. \ref{fig:2}b were subsequently filtered by a low-pass filter with a bandwidth corresponding to the cut-off frequency (10 Hz) of the electrical circuit used for the resistance measurements. 

%%%%%%%%%%%%%%%%%%%%%%%%%%%%%%%%%%%%%%%%%%%%%%%%%%%%%%%%%%
\section{Results and discussion}
\begin{figure}
\begin{center}
\begin{tabular}{c}
\includegraphics[width=8.2cm]{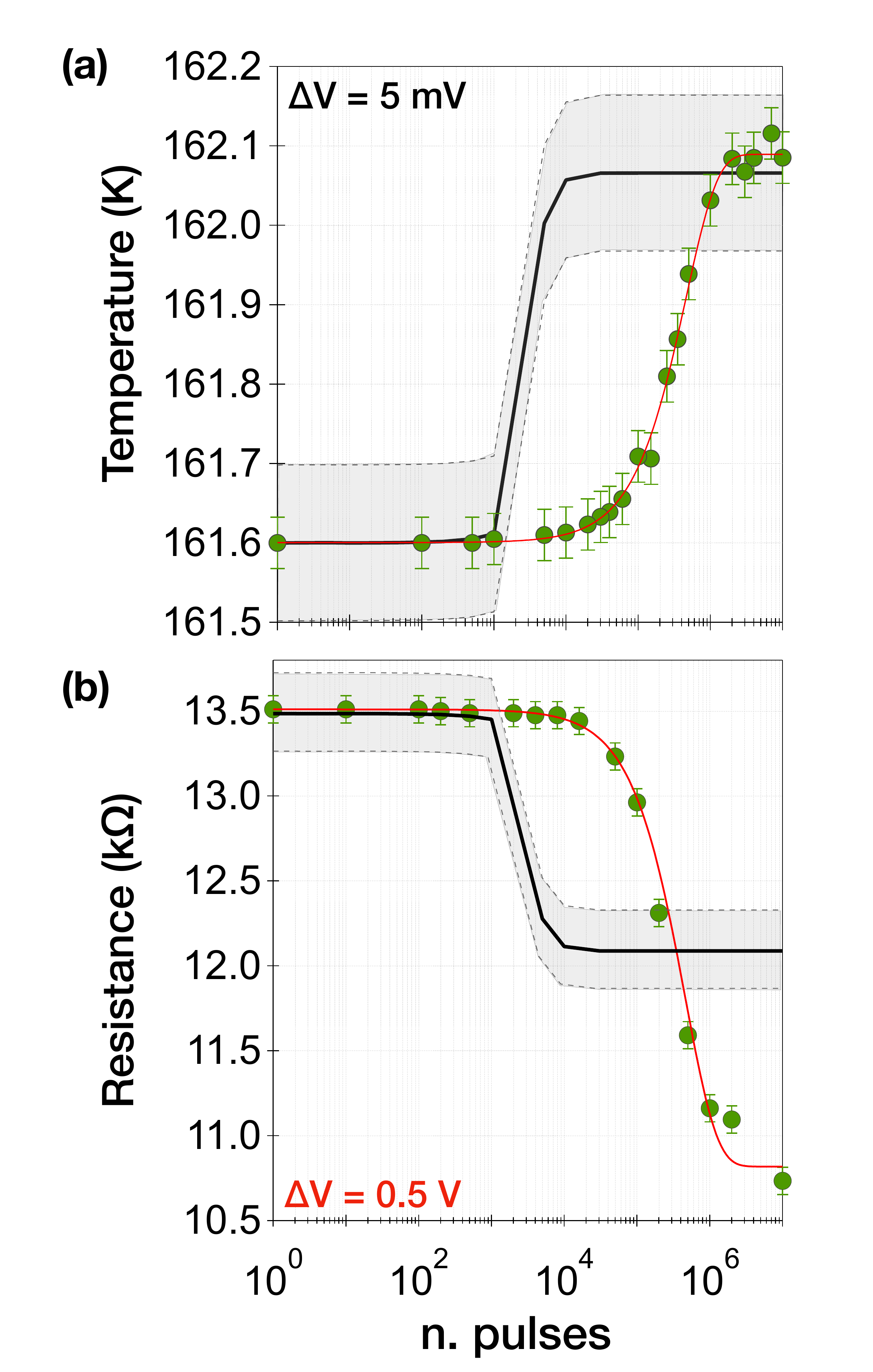}
\end{tabular}
\end{center}
\caption 
{ \label{fig:3}
\textbf{Comparison between experimental and numerical results.} \textbf{(a)} Maximum variation of the sample's temperature induced by the laser pulses as function of the total number of pulses (green circles). The applied voltage is $\Delta V$=5 mV. The solid black line represents the simulated data where the maximum temperature has been extracted for each finite number of pulses. The grey area shows the error interval associated to the model assumptions and numerical approximations. The red solid line represents the exponential fit to the experimental data. \textbf{(b)} Maximum laser-induced drop of the device resistance when a constant voltage bias $\Delta V$=0.5 \si{\volt} is applied across the Au/Ti electrodes (green circles). The solid black line represents the outcome of numerical simulations, from which the expected resistance drop is retrieved from the local effective heating for each finite number of pulses. The dissipated power related to the Joule heating has been added to the power dissipated by the laser excitation. The grey areas show the error interval associated to the model assumptions and numerical approximations. The solid red line represents the exponential fit to the experimental data.} 
\end{figure}

The comparison between the experimental results and the numerical simulations is shown in Fig. \ref{fig:3}a. The applied voltage for the resistance measurements is $\Delta V$=5 \si{m\volt}, i.e. much smaller than the threshold $\Delta V_{th}\simeq$2.5 V necessary to obtain the resistive switching at the working temperature. The experimental points (green circles) represent the maximum effective temperature increase, estimated by the volatile resistance drop described in Sec \ref{sec:PI_res_measurements}, as a function of the number of excitation pulses. We stress that each experimental point is obtained after a complete thermal cycle, necessary to restore the initial resistance of the device (see Sec. \ref{sec:device}). The data show that no significant temperature increase is detectable below $\sim 10^{4}$ pulses. Above this threshold the temperature increases and saturates after approximately $10^{6}$ pulses. The temperature achieved in the saturation regime, i.e. when the excitation average power matches the thermal losses towards the substrate, is $\sim$ 162.06 \si{\kelvin}. 

On general grounds, the dynamics of the V$_2$O$_3$ temperature $T(t)$ can be described in terms of a simple lumped-element model:
\begin{equation}
\label{lumped}
C_t\frac{d(T-T_0)}{dt}=W-\frac{1}{R_{th}}(T-T_0)
\end{equation}
where $T_0$=161.6 K is the initial temperature, $C_t$ represents the total heat capacity of the device, $W$ is the total absorbed power, $R_{th}$ is a constant with the dimensions of a thermal resistance and effectively accounts for the total heat dissipation. Independently of the specific values of the effective coefficients entering in the lumped-element model, the general solution of Eq. \ref{lumped} is the exponential function $T(t)$=$T_0$+$\delta T_{sat}$(1-$e^{-t/\tau}$), where $\tau$=$R_{th}C_t$ is the time constant of the heating up process and $\delta T_{sat}$=$R_{th}W$ is the temperature variation at saturation. This exponential function can be used to fit the data and retrieve the experimental time constant of the system. The red solid line in Fig. \ref{fig:3}a is the exponential fit, which provides $\tau$=18.5$\pm$0.8 s corresponding to (4.6$\pm$0.2)$\times$10$^5$ pulses. 

The local effective temperature retrieved by the resistance drop is compared to the results of calculations of the full thermal problem, as described in Sec. \ref{sec:thermal}. The black solid line in Fig. \ref{fig:3}a represents the saturation temperature increase obtained by finite-element calculations, when an average power corresponding to that of the pulse train exciting the device is considered. The calculated saturation temperature increase, i.e. $\delta T_{sat}$=0.5 K perfectly matches the experimental value. Considering that $\delta T_{sat}$ depends only on the adsorbed power $W$ and on $R_{th}$, we can conclude that the finite element calculations perfectly account for the heat dissipation, which is ultimately regulated by the heat diffusion throughout the V$_2$O$_3$ film and the substrate. 

The difference between the calculated and measured time constants of the exponential heating dynamics can be attributed to an underestimation of the total heat capacity $C_t$. Throughout the calculations of the thermal problem, we assumed perfect electron-phonon thermalization and we have thus considered the V$_2$O$_3$ temperature-dependent total specific heat \cite{Keer1976}. Nonetheless, the absorption of each single light pulse is mediated by the electronic population which undergoes a large temperature increase. The effective electronic temperature  dynamically overcomes the Mott insulator-to-metal transition temperature $T_{IMT}$, which is accompanied by a divergence of the electronic effective mass $m^*$. Considering that the electronic specific heat of an electron gas $c_{el}\propto {m^*}^{3/2}$, the dynamical crossing of  $T_{IMT}$ is naturally accompanied by a large increase of $c_{el}$ \cite{Keer1976}, which is characterized by a narrow cusp centered at $T_{IMT}$ (see Tab. \ref{tab:1}). In the vicinity of the Mott transition, the electron gas thus behaves like a reservoir capable of absorbing a large quantity of energy but limiting the impulsive increase of the effective temperature of the electron-phonon system. Our experimental results are compatible with $c_{el}$=(30$\pm$6)$\times c_p(T_{IMT})\big|_{V_2O_3}$, which suggests an effective mass increase of 10, perfectly compatible with what recently observed in VO$_2$ \cite{Lee2017} and attributed to the strong electronic correlations.

Once assessed the validity of the thermal model to evaluate the laser induced heating in the saturation regime, we can compare the expected thermally induced resistance drop to what is actually measured when the laser pulses are combined to a voltage that is no longer negligible with respect to $\Delta V_{th}$. In Fig. \ref{fig:3}b we report the resistance drop, measured following the procedure previously described, with $\Delta V$=0.5 V applied voltage. In this case, although the resistance drop is described by an exponential function (solid red line) with the same time constant ($\tau$=18.7$\pm$1.6 s), the value achieved in the saturation regime ($\delta$R$_{sat} \simeq$2.6 k$\Omega$) is twice the value measured ($\delta$R$_{sat} \simeq$1.3 k$\Omega$) for $\Delta V$=5 mV.

The resistance drop measured under the applied bias $\Delta V$=0.5 V is compared to the value expected by considering the laser-induced heating (the parameters of the thermal calculations are the same than the previous low-voltage case) and also including the effect of Joule heating induced by the current flowing throughout the V$_{2}$O$_{3}$ bridge. We stress that the voltage applied through the device is smaller than threshold value necessary for the resistive switching. Therefore, in this regime no metallic filaments\cite{Kalcheim2020}, possibly responsible for large local Joule heating, are created and the system can be treated as homogeneous. The additional dissipated power can be thus written as:
\begin{equation}
\Delta P_J =\frac{\Delta V\times I}{swd}
\label{e_dissipation}
\end{equation}
where $\Delta P_J$ is the power density dissipated by the electrical current $I$ flowing through the electrodes and $swd$ is the device effective volume (see Fig. \ref{fig:1}a). $\Delta P_J$ is added to the total power density $P(r,z)$, provided by the pulse train, in Eq \ref{eq:1}. The black solid line in Fig. \ref{fig:3}b represents the result of the numerical simulation converted in the expected resistance drop by means of the resistance vs temperature calibration curve (see Fig. \ref{fig:1}b). We note that the Joule heating induced by the current is of the order of 2$\times 10^{-5}$ \si{W} and is negligible as compared to the laser power, which amounts to $\sim 10^{-2}$ \si{W}. As a conclusion, the additional Joule heating does not account for the measured resistivity drop in the stationary regime, which is twice than that expected from heating alone. 

The present results can be rationalized on the basis of recent theoretical work\cite{Mazza2016}, which showed that in the insulator-metal coexistence region, the application of a significant electric field is able to induce the non-thermal weakening of the Mott insulating phase via the formation of new metallic states at the Fermi level with no counterpart at equilibrium. This mechanism is in sharp contrast with the conventional Zener tunneling mechanism that is usually invoked to describe the resistive switching dynamics. Interestingly, the state variable that characterizes the transition and controls the free energy of the system is the orbital polarization, defined as $p$=$n_{e_g^{\pi}}$-$n_{a_{1g}}$, where $n_{e_g^{\pi}}$ is the occupation of the lowest energy V-3$d$ $e_g^{\pi}$ orbitals and $n_{a_{1g}}$ is the occupation of the V-3$d$ $a_{1g}$ orbitals. The Mott insulating phase is characterized by $p$=2, whereas the application of an electric field induces the formation of a metallic state with $p$=1.2-1.6 \cite{Mazza2016}. As recently shown\cite{Ronchi2019}, the excitation with infrared light pulses with 1.55 eV photon energy can induce a further orbital polarization decrease of the order of $\delta p \simeq$-10$^{-3}-10^{-2}$, which in turns leads to a change of 1.3-13 meV of the difference between the free energies of the insulating and metallic phases. The photo-induced change of the free energy difference largely overcomes the thermal effect, i.e. $k_B \delta T \simeq$50 $\mu$eV, thus inducing the growth of already existing metallic domains in the insulator-metal coexistence region. The combination of below-threshold voltage and the excitation with infrared light pulses capable of modifying the occupation of vanadium orbitals thus weakens the insulating state well beyond what can be ascribed to thermal effects. The scenario emerging from our results can be understood also within the resistor network model \cite{Stoliar2013}, which is commonly adopted to describe the insulator-to-metal switching dynamics. In this model, the physical system is divided in nanometric cells representing small regions, which are either in the insulating (high-resistance) or metallic (low-resistance) states. The combined action of voltage and laser excitation leads to a non-thermal increase of the number of the metallic cells, thus triggering the observed volatile resistance drop. Working with below-threshold voltage and with weak laser excitation avoids the creation of connected filamentary metallic paths which are responsible for the resistive switching process observed above $\Delta V_{th}$.

\section{Conclusions}\label{Conclusions}
In conclusion, we investigated the simultaneous action of a train of infrared light pulses and an external voltage on a Mottronic device based on V$_2$O$_3$, which undergoes an IMT at $T_{IMT}\simeq$170 K. When the device is in the insulator-metal coexisting region ($T\simeq$160 K) and a 0.5 V voltage is applied across the device, the measured resistance drop induced by a pulse number exceeding 10$^6$ is twice than that expected from the simple local heating of the device. Our results suggest that although the 0.5 V applied voltage is below the threshold ($\Delta V_{th}$=2.5 V) necessary for inducing the complete and irreversible resistive switch, it brings the insulating phase close to the insulator-to-metal instability. The simultaneous weak excitation with a sufficient number of infrared pulses modifies the occupation of the vanadium 3$d$ orbitals enough to trigger the non-thermal growth and proliferation of already existing metallic nodes, well beyond what can be ascribed to the total local heating induced by the laser/voltage combination. The present results also call for the development of time-resolved microscopy techniques to investigate the real time dynamics of the metallicity in Mottronics and resistive switching devices subject to the simultaneous application of electric fields and light pulses.    

\begin{acknowledgments}
C.G., A.R. and P.F. acknowledge Andrea Tognazzi for the support in the development of the experimental set-up. C.G. and A.R. acknowledge Dr. Alessandro Bau' and Prof. Vittorio Ferrari (Information Engeenering Dept., Universit\`a degli Studi di Brescia) for the support given during the wirebonding of the device. We thank Frederik Ceyssens for his help with the fabrication of electrical contacts on the samples.
C.G., A.R., P.F. acknowledge financial support from MIUR through the PRIN 2017 program (Prot. 20172H2SC4\_005). G.F., and C.G. acknowledge support from Universit\`a Cattolica del Sacro Cuore through D.1, D.2.2, and D.3.1 grants. 
F.B. acknowledges financial support from Universit\'e de Lyon in the frame of the IDEXLYON Project-Programme Investissements d’Avenir (ANR-16-IDEX-0005) and from Université Claude Bernard Lyon 1 thorugh the BQR Accueil EC 2019 grant. P.H., M.M., and J.-P.L. acknowledge support from EU-H2020-ICT-2015 PHRESCO Project, Grant agreement No. 688579. P.H. acknowledges support from Becas Chile-CONICYT. %F.P. acknowledge the financial support from UniTS through the FRA 2016 Research Programme. 
M. G. acknowledges financial support from the CNR Joint Laboratories program 2019-2021.

\end{acknowledgments}


\begin{thebibliography}{10}

\bibitem{Tokura2017}
Yoshinori Tokura, Masashi Kawasaki, and Naoto Nagaosa.
\newblock Emergent functions of quantum materials.
\newblock {\em Nature Physics}, 13(11):1056--1068, September 2017.

\bibitem{Yang2011}
Zheng Yang, Changhyun Ko, and Shriram Ramanathan.
\newblock Oxide electronics utilizing ultrafast metal-insulator transitions.
\newblock {\em Annual Review of Materials Research}, 41(1):337--367, 2011.

\bibitem{Zhang2014}
J.~Zhang and R.D. Averitt.
\newblock Dynamics and control in complex transition metal oxides.
\newblock {\em Annual Review of Materials Research}, 44(1):19--43, 2014.

\bibitem{Basov2017}
D.~N. Basov, R.~D. Averitt, and D.~Hsieh.
\newblock Towards properties on demand in quantum materials.
\newblock {\em Nature Materials}, 16(11):1077--1088, November 2017.

\bibitem{Zhou2015}
Y.~{Zhou} and S.~{Ramanathan}.
\newblock Mott memory and neuromorphic devices.
\newblock {\em Proceedings of the IEEE}, 103(8):1289--1310, 2015.

\bibitem{Janod2015}
Etienne Janod, Julien Tranchant, Benoit Corraze, Madec Querr\'e, Pablo Stoliar,
  Marcelo Rozenberg, Tristan Cren, Dimitri Roditchev, Vinh~Ta Phuoc,
  Marie-Paule Besland, and Laurent Cario.
\newblock Resistive switching in mott insulators and correlated systems.
\newblock {\em Advanced Functional Materials}, 25(40):6287--6305, 2015.

\bibitem{delValle2018}
Javier del Valle, Juan~Gabriel Ram\'irez, Marcelo~J. Rozenberg, and Ivan~K.
  Schuller.
\newblock Challenges in materials and devices for resistive-switching-based
  neuromorphic computing.
\newblock {\em Journal of Applied Physics}, 124(21):211101, 2018.

\bibitem{Salev2019}
Pavel Salev, Javier del Valle, Yoav Kalcheim, and Ivan~K. Schuller.
\newblock Giant nonvolatile resistive switching in a mott oxide and
  ferroelectric hybrid.
\newblock {\em Proceedings of the National Academy of Sciences},
  116(18):8798--8802, 2019.

\bibitem{Kuwamoto1980}
H.~Kuwamoto, J.~M. Honig, and J.~Appel.
\newblock Electrical properties of the \ce{(V_{1-x}Cr_x)_2O_3} system.
\newblock {\em Phys. Rev. B}, 22:2626--2636, Sep 1980.

\bibitem{Homm2015}
P.~Homm, L.~Dillemans, M.~Menghini, B.~Van~Bilzen, P.~Bakalov, C.-Y. Su,
  R.~Lieten, M.~Houssa, D.~Nasr~Esfahani, L.~Covaci, F.~M. Peeters, J.~W. Seo,
  and J.-P. Locquet.
\newblock {}collapse of the low temperature insulating state in cr-doped
  \ce{V2O3} thin films.

\bibitem{delvalle2017}
Javier del Valle, Yoav Kalcheim, Juan Trastoy, Aliaksei Charnukha, Dimitri~N.
  Basov, and Ivan~K. Schuller.
\newblock Electrically induced multiple metal-insulator transitions in oxide
  nanodevices.
\newblock {\em Phys. Rev. Applied}, 8:054041, Nov 2017.

\bibitem{Ronchi2019}
A.~Ronchi, P.~Homm, M.~Menghini, P.~Franceschini, F.~Maccherozzi, F.~Banfi,
  G.~Ferrini, F.~Cilento, F.~Parmigiani, S.~S. Dhesi, M.~Fabrizio, J.-P.
  Locquet, and C.~Giannetti.
\newblock {Early-stage dynamics of metallic droplets embedded in the
  nanotextured Mott insulating phase of {${\mathrm{V}}_{2}{\mathrm{O}}_{3}$}}.
\newblock {\em Phys. Rev. B}, 100:075111, Aug 2019.

\bibitem{Mcleod2016}
A.~S. McLeod, E.~van Heumen, J.~G. Ramirez, S.~Wang, T.~Saerbeck, S.~Guenon,
  M.~Goldflam, L.~Anderegg, P.~Kelly, A.~Mueller, M.~K. Liu, Ivan~K. Schuller,
  and D.~N. Basov.
\newblock Nanotextured phase coexistence in the correlated insulator \ce{V2O3}.
\newblock {\em Nature Physics}, 13(1):80--86, September 2016.

\bibitem{McWhan1969}
D.~B. McWhan, T.~M. Rice, and J.~P. Remeika.
\newblock Mott transition in {Cr}-doped \ce{V2O3}.
\newblock {\em Phys. Rev. Lett.}, 23:1384--1387, Dec 1969.

\bibitem{Jayaraman1970}
A.~Jayaraman, D.~B. McWhan, J.~P. Remeika, and P.~D. Dernier.
\newblock Critical behavior of the {Mott} transition in cr-doped \ce{V2O3}.
\newblock {\em Phys. Rev. B}, 2:3751--3756, Nov 1970.

\bibitem{Valmianski2017}
I.~Valmianski, Juan~Gabriel Ramirez, C.~Urban, X.~Batlle, and Ivan~K. Schuller.
\newblock Deviation from bulk in the pressure-temperature phase diagram of
  \ce{V2O3} thin films.
\newblock {\em Phys. Rev. B}, 95:155132, Apr 2017.

\bibitem{Limelette89}
P.~Limelette, A.~Georges, D.~J{\'e}rome, P.~Wzietek, P.~Metcalf, and J.~M.
  Honig.
\newblock Universality and critical behavior at the {Mott} transition.
\newblock {\em Science}, 302(5642):89--92, 2003.

\bibitem{Dillemans2014}
L.~Dillemans, T.~Smets, R.~R. Lieten, M.~Menghini, C.-Y. Su, and J.-P. Locquet.
\newblock Evidence of the metal-insulator transition in ultrathin unstrained
  \ce{V2O3} thin films.
\newblock {\em Applied Physics Letters}, 104(7):071902, 2014.

\bibitem{Lee2018}
D.~Lee, B.~Chung, Y.~Shi, G.-Y. Kim, N.~Campbell, F.~Xue, K.~Song, S.-Y. Choi,
  J.~P. Podkaminer, T.~H. Kim, P.~J. Ryan, J.-W. Kim, T.~R. Paudel, J.-H. Kang,
  J.~W. Spinuzzi, D.~A. Tenne, E.~Y. Tsymbal, M.~S. Rzchowski, L.~Q. Chen,
  J.~Lee, and C.~B. Eom.
\newblock Isostructural metal-insulator transition in \ce{VO2}.
\newblock {\em Science}, 362(6418):1037--1040, 2018.

\bibitem{Kalcheim2019}
Yoav Kalcheim, Nikita Butakov, Nicolas~M. Vargas, Min-Han Lee, Javier del
  Valle, Juan Trastoy, Pavel Salev, Jon Schuller, and Ivan~K. Schuller.
\newblock Robust coupling between structural and electronic transitions in a
  {Mott} material.
\newblock {\em Phys. Rev. Lett.}, 122:057601, Feb 2019.

\bibitem{Stoliar2013}
Pablo Stoliar, Laurent Cario, Etiene Janod, Benoit Corraze, Catherine
  Guillot-Deudon, Sabrina Salmon-Bourmand, Vincent Guiot, Julien Tranchant, and
  Marcelo Rozenberg.
\newblock Universal electric-field-driven resistive transition in narrow-gap
  {Mott} insulators.
\newblock {\em Advanced Materials}, 25(23):3222--3226, 2013.

\bibitem{Guenon2013}
S.~Gu{\'{e}}non, S.~Scharinger, Siming Wang, J.~G. Ram{\'{\i}}rez, D.~Koelle,
  R.~Kleiner, and Ivan~K. Schuller.
\newblock Electrical breakdown in a \ce{V2O3} device at the insulator-to-metal
  transition.
\newblock {\em {EPL} (Europhysics Letters)}, 101(5):57003, mar 2013.

\bibitem{Mazza2016}
G.~Mazza, A.~Amaricci, M.~Capone, and M.~Fabrizio.
\newblock Field-driven {Mott} gap collapse and resistive switch in correlated
  insulators.
\newblock {\em Phys. Rev. Lett.}, 117:176401, Oct 2016.

\bibitem{DelValle2019}
Javier del Valle, Pavel Salev, Federico Tesler, Nicolás~M. Vargas, Yoav
  Kalcheim, Paul Wang, Juan Trastoy, Min-Han Lee, George Kassabian,
  Juan~Gabriel Ramírez, Marcelo~J. Rozenberg, and Ivan~K. Schuller.
\newblock Subthreshold firing in {Mott} nanodevices.
\newblock {\em Nature}, 569(7756):388--392, May 2019.

\bibitem{Kalcheim2020}
Yoav Kalcheim, Alberto Camjayi, Javier del Valle, Pavel Salev, Marcelo
  Rozenberg, and Ivan~K Schuller.
\newblock {Non-thermal resistive switching in {Mott} insulator nanowires}.
\newblock {\em Nature Communications}, 11(1):2985, 2020.

\bibitem{Mansart2010}
B.~Mansart, D.~Boschetto, S.~Sauvage, A.~Rousse, and M.~Marsi.
\newblock Mott transition in cr-doped \ce{V2O3} studied by ultrafast
  reflectivity: Electron correlation effects on the transient response.
\newblock {\em {EPL} (Europhysics Letters)}, 92(3):37007, nov 2010.

\bibitem{Liu2011}
M.~K. Liu, B.~Pardo, J.~Zhang, M.~M. Qazilbash, Sun~Jin Yun, Z.~Fei, Jun-Hwan
  Shin, Hyun-Tak Kim, D.~N. Basov, and R.~D. Averitt.
\newblock Photoinduced phase transitions by time-resolved far-infrared
  spectroscopy in \ce{V2O3}.
\newblock {\em Phys. Rev. Lett.}, 107:066403, Aug 2011.

\bibitem{Abreu2015}
Elsa Abreu, Siming Wang, Juan~Gabriel Ram\'{\i}rez, Mengkun Liu, Jingdi Zhang,
  Kun Geng, Ivan~K. Schuller, and Richard~D. Averitt.
\newblock Dynamic conductivity scaling in photoexcited \ce{V2O3} thin films.
\newblock {\em Phys. Rev. B}, 92:085130, Aug 2015.

\bibitem{Morrison2014}
Vance~R. Morrison, Robert.~P. Chatelain, Kunal~L. Tiwari, Ali Hendaoui, Andrew
  Bruh{\'a}cs, Mohamed Chaker, and Bradley~J. Siwick.
\newblock A photoinduced metal-like phase of monoclinic \ce{VO2} revealed by
  ultrafast electron diffraction.
\newblock {\em Science}, 346(6208):445--448, 2014.

\bibitem{Abreu2017}
Elsa Abreu, Stephanie~N. Gilbert~Corder, Sun~Jin Yun, Siming Wang, Juan~Gabriel
  Ram\'{\i}rez, Kevin West, Jingdi Zhang, Salinporn Kittiwatanakul, Ivan~K.
  Schuller, Jiwei Lu, Stuart~A. Wolf, Hyun-Tak Kim, Mengkun Liu, and Richard~D.
  Averitt.
\newblock Ultrafast electron-lattice coupling dynamics in \ce{VO2} and
  \ce{V2O3} thin films.
\newblock {\em Phys. Rev. B}, 96:094309, Sep 2017.

\bibitem{Lantz2017}
G.~Lantz, B.~Mansart, D.~Grieger, D.~Boschetto, N.~Nilforoushan,
  E.~Papalazarou, N.~Moisan, L.~Perfetti, V.~L.~R. Jacques, D.~Le~Bolloc'h,
  C.~Laulhé, S.~Ravy, J-P Rueff, T.~E. Glover, M.~P. Hertlein, Z.~Hussain,
  S.~Song, M.~Chollet, M.~Fabrizio, and M.~Marsi.
\newblock Ultrafast evolution and transient phases of a prototype
  out-of-equilibrium {Mott}–{Hubbard} material.
\newblock {\em Nature Communications}, 8(1):13917, January 2017.

\bibitem{Otto2019}
Martin~R. Otto, Laurent~P. Ren{\'e}~de Cotret, David~A. Valverde-Chavez,
  Kunal~L. Tiwari, Nicolas {\'E}mond, Mohamed Chaker, David~G. Cooke, and
  Bradley~J. Siwick.
\newblock How optical excitation controls the structure and properties of
  vanadium dioxide.
\newblock {\em Proceedings of the National Academy of Sciences},
  116(2):450--455, 2019.

\bibitem{Giorgianni2019}
Flavio Giorgianni, Joe Sakai, and Stefano Lupi.
\newblock Overcoming the thermal regime for the electric-field driven {Mott}
  transition in vanadium sesquioxide.
\newblock {\em Nature Communications}, 10(1):1159, March 2019.

\bibitem{Lupi2010}
S~Lupi, L~Baldassarre, B~Mansart, A~Perucchi, A~Barinov, P~Dudin,
  E~Papalazarou, F~Rodolakis, J~P. Rueff, J~P. Iti{\'{e}}, S~Ravy, D~Nicoletti,
  P~Postorino, P~Hansmann, N~Parragh, A~Toschi, T~Saha-Dasgupta, O~K Andersen,
  G~Sangiovanni, K~Held, and M~Marsi.
\newblock {A microscopic view on the Mott transition in chromium-doped
  \ce{V2O3}}.
\newblock {\em Nature Communications}, 1:105, 2010.

\bibitem{Sandri2015}
Matteo Sandri and Michele Fabrizio.
\newblock Nonequilibrium gap collapse near a first-order {Mott} transition.
\newblock {\em Phys. Rev. B}, 91:115102, Mar 2015.

\bibitem{Qazilbash2008}
M.~M. Qazilbash, A.~A. Schafgans, K.~S. Burch, S.~J. Yun, B.~G. Chae, B.~J.
  Kim, H.~T. Kim, and D.~N. Basov.
\newblock Electrodynamics of the vanadium oxides \ce{VO2} and \ce{V2O3}.
\newblock {\em Phys. Rev. B}, 77:115121, Mar 2008.

\bibitem{Andreev1978}
V.~N. Andreev, F.~A. Chudnovskii, A.~V. Petrov, and E.~I. Terukov.
\newblock Thermal conductivity of \ce{VO2}, \ce{V3O5}, and \ce{V2O3}.
\newblock {\em physica status solidi (a)}, 48(2):K153--K156, 1978.

\bibitem{Keer1976}
H.V. Keer, D.L. Dickerson, H.~Kuwamoto, H.L.C. Barros, and J.M. Honig.
\newblock Heat capacity of pure and doped \ce{V2O3} single crystals.
\newblock {\em Journal of Solid State Chemistry}, 19(1):95 -- 102, 1976.

\bibitem{Weber2001}
M.~J. Weber.
\newblock {\em Handbook of Optical Materials}.
\newblock CRC Press, Boca Raton, FL, 2001.

\bibitem{Caddeo2017}
Claudia Caddeo, Claudio Melis, Andrea Ronchi, Claudio Giannetti, Gabriele
  Ferrini, Riccardo Rurali, Luciano Colombo, and Francesco Banfi.
\newblock Thermal boundary resistance from transient nanocalorimetry: A
  multiscale modeling approach.
\newblock {\em Phys. Rev. B}, 95:085306, Feb 2017.

\bibitem{Malitson1962}
Irving~H. Malitson.
\newblock Refraction and dispersion of synthetic sapphire.
\newblock {\em J. Opt. Soc. Am.}, 52(12):1377--1379, Dec 1962.

\bibitem{Hamaoui2019}
Georges Hamaoui, Nicolas Horny, Cindy~Lorena Gomez-Heredia, Jorge~Andres
  Ramirez-Rincon, Jose Ordonez-Miranda, Corinne Champeaux, Frederic
  Dumas-Bouchiat, Juan~Jose Alvarado-Gil, Younes Ezzahri, Karl Joulain, and
  Mihai Chirtoc.
\newblock {Thermophysical characterisation of \ce{VO2} thin films hysteresis
  and its application in thermal rectification}.
\newblock {\em Scientific reports}, 9(1):8728, jun 2019.

\bibitem{Giannetti2009}
C.~{Giannetti}, F.~{Banfi}, D.~{Nardi}, G.~{Ferrini}, and F.~{Parmigiani}.
\newblock Ultrafast laser pulses to detect and generate fast thermomechanical
  transients in matter.
\newblock {\em IEEE Photonics Journal}, 1(1):21--32, 2009.

\bibitem{Banfi2010}
F.~Banfi, F.~Pressacco, B.~Revaz, C.~Giannetti, D.~Nardi, G.~Ferrini, and
  F.~Parmigiani.
\newblock Ab initio thermodynamics calculation of all-optical time-resolved
  calorimetry of nanosize systems: Evidence of nanosecond decoupling of
  electron and phonon temperatures.
\newblock {\em Phys. Rev. B}, 81:155426, Apr 2010.

\bibitem{Lee2017}
Sangwook Lee, Kedar Hippalgaonkar, Fan Yang, Jiawang Hong, Changhyun Ko, Joonki
  Suh, Kai Liu, Kevin Wang, Jeffrey~J. Urban, Xiang Zhang, Chris Dames, Sean~A.
  Hartnoll, Olivier Delaire, and Junqiao Wu.
\newblock Anomalously low electronic thermal conductivity in metallic vanadium
  dioxide.
\newblock {\em Science}, 355(6323):371--374, 2017.

\end{thebibliography}
\end{document}